\documentclass[10pt]{iopart}

\usepackage{siunitx}
\usepackage{graphicx}
\begin{document}

\title{Patterning of diamond with 10 nm resolution by electron-beam-induced etching}

\author[1]{Vasilis Dergianlis, Martin Geller, Dennis Oing, Nicolas Wöhrl and Axel Lorke}

\address{University of Duisburg-Essen, Faculty of Physics and CENIDE, D-47057 Duisburg, Germany}
\ead{vasilis.dergianlis@uni-duisburg-essen.de}
\vspace{10pt}
\begin{indented}
\item[]March 2019
\end{indented}

\begin{abstract}
We report on mask-less, high resolution etching of diamond surfaces, featuring sizes down to 10 nm. We use a scanning electron microscope (SEM) together with water vapor, which was injected by a needle directly onto the sample surface. Using this versatile and low-damage technique, trenches with different depths were etched. Cross sections of each trench were obtained by focused ion beam milling and used to calculate the achieved aspect ratios. The developed technique opens up the possibility of mask- and resist-less patterning of diamond for nano-optical and electronic applications. 
\end{abstract}

%
\vspace{2pc}


\section{Introduction}

Diamond is a material that possesses unique and attractive properties including ultrahigh hardness, chemical stability and mechanical strength, while it is highly transparent and an exceptional thermal conductor\cite{Loncar2013}. Furthermore, diamond is a wide-bandgap semiconductor with working temperatures up to 500\si{\celsius}. It has a bandgap of 5.47 eV at room temperature and its surface can be functionalized to exhibit negative electron affinity. Due to these physical properties in combination with the capability of hosting nitrogen vacancy luminescence centers (NV), diamond is one of the most promising candidates as a platform for next generation sensing, nanophotonic and quantum information devices\cite{Aharonovich2014, Loncar2013}.  

The processing of diamond for its implementation in functional devices is, however, extremely challenging due to the very same physical properties that render it an excellent candidate for these applications. Currently, high resolution diamond structuring requires arduous masking techniques and ion bombardment\cite{Bayn2011a, Taniguchi2002} or high power laser ablation\cite{Lehmann2014}, which often cause damage and material re-deposition artifacts\cite{Martin2014, Bayn2011}
At present, the prevailing method for the mask-less patterning of diamond is Focused Ion Beam (FIB) milling, which however limits the fabrication process due to extended surface damage that leads to graphitization, re-deposition of carbon, ion implantation and low resolution. \cite{Uzan-Saguy1995, Bayn2011a}

As a less-destructive patterning technique of diamond layers, gas-assisted Electron Beam Induced Etching (EBIE)\cite{Randolph2006, Hoffmann2005} was proposed, which is a well-known technique being used for high resolution patterning of semiconductors  such as GaAs\cite{Ganczarczyk2011}  and Ge\cite{Gokdeniz2015} as well as graphene\cite{Sommer2014}. Gas assisted EBIE to pattern diamond was first proposed by Taniguchi et al.\cite{Taniguchi1996} and we show here that is an alternative, low-damage method that opens the way for mask-less patterning of diamond for nanoscale optoelectronic applications without the need of the FIB. This method is based on the combination of an electron beam from a Scanning Electron Microscope (SEM) and \textit{in-situ} exposure to a suitable gaseous etchant. Common etchants are oxygen\cite{Taniguchi1996}, hydrogen\cite{Taniguchi1997} and water vapor\cite{Niitsuma2006}. 
The working principle behind this method is that the electron beam ionizes the gas molecules, which then create volatile compounds with the carbon atoms on the diamond surface\cite{Martin2013, Martin2015, Taniguchi1997}. After the first reports on the method\cite{Taniguchi1996, Taniguchi1997}, further work has been published, studying the effect of different gases and pressure \cite{Niitsuma2006, Martin2015a} on the etching process. More recently, the patterns that are being created on the surface of diamond during etching were studied in correlation with the gases used in each case \cite{Martin2015, Bishop2018}. However, the method was not studied until now in terms of high resolution in the range of nanometers for direct 3D, mask-less device nano-fabrication on the surface of diamond. The best resolution reported so far on single crystalline diamond was $\sim{100}$ nm \cite{Martin2014}.  In this work, we report the mask-less and high resolution patterning of hydrogen-terminated single crystalline diamond samples using water vapor assisted EBIE featuring sizes down to 10 nm.

\section{Sample synthesis and experimental setup}
\subsection{Sample synthesis}

\begin{figure*}[h]
	\centering
	\includegraphics[width=\textwidth]{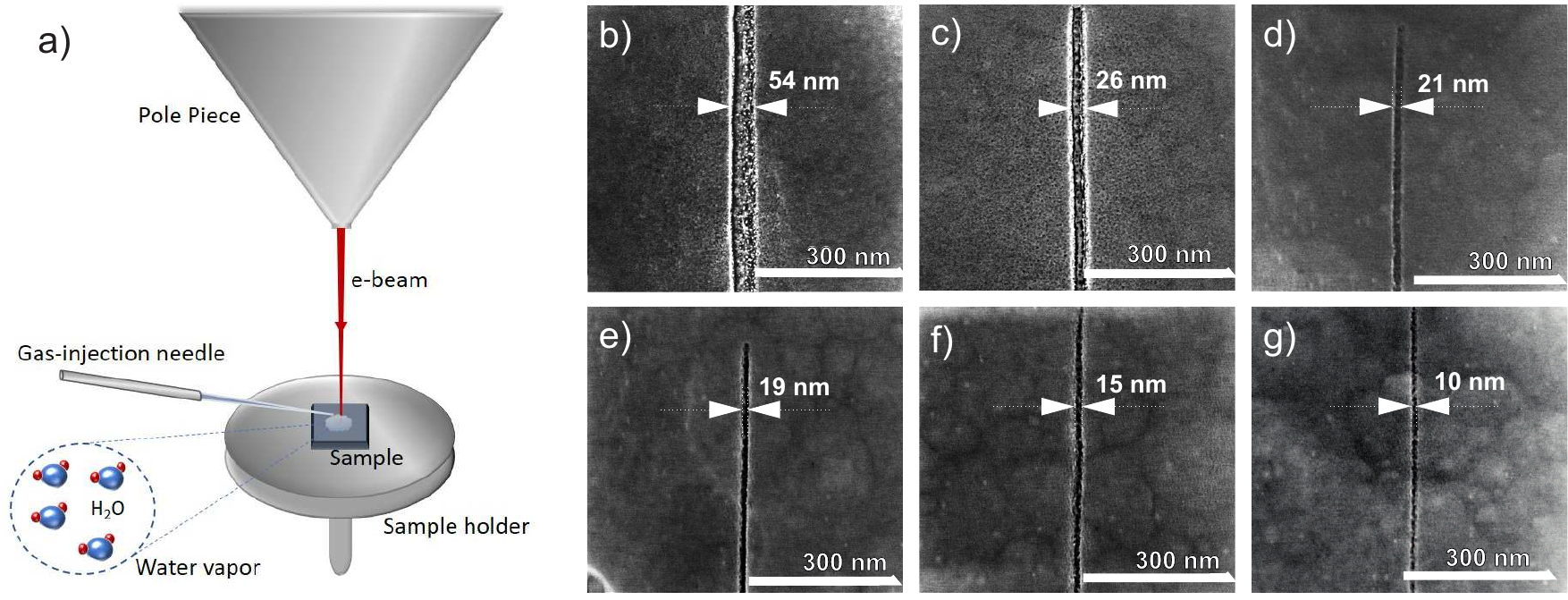}
	\caption{a) Experimental setup. The sample is mounted on the sample holder, while the gas-injection needle is releasing water vapor above the diamond surface. The electron beam coming out of the SEM's pole piece ionizes the water molecules through impact ionization and the resulting ions create volatile carbon compounds at the diamond surface. (b)-(g) SEM images of etched trenches for the same dose of approximately 10 nC/$\mu$m and different e-beam parameters b) 10 kV and 21 pA c) 10 kV and 43 pA, d) 30 kV and 43 pA, e) 20 kV and 0.34 nA, f) 20 kV and 0.69 nA and g) 30 kV and 0.69 nA.}
	\label{Exp_setup_HR}
\end{figure*}

The samples were grown in a microwave plasma-assisted CVD reactor based on a 2.45 GHz IPLAS CYRANNUS I-6'' plasma source\cite{Aschermann1997}. The single-crystal, \{100\}-oriented diamond samples were placed on the substrate holder and the chamber was pumped down to $<10^{-7}$  mbar. Then a hydrogen plasma at 200 mbar, with a microwave power of 2.6 kW and a gas flow of 400 sccm for 45 min was used to clean the sample and sample holder prior to deposition. The deposition started at 120 mbar chamber pressure, with 1.26 kW microwave power. The $H_{2}$ flow was at 500 sccm with a purity of 7.0, while for methane $CH_{4}$ was 57 sccm with a purity of 9.0 and a substrate temperature of 810\si{\celsius}. Deposition took 6 hours, resulting in an approximately 17 $\mu$m thin film with \{100\} orientation. These surfaces were used for the etching experiments as grown.
\subsection{Experimental setup}
For the EBIE, a FEI Helios 600 NanoLab DualBeam setup was used, which integrates an electron and an ion beam. This system implements a gas injection system (GIS) for enhanced etching and material removal, as well as metal and insulator deposition. For the purpose of our study, only the electron beam was utilized in combination with simultaneous injection of water vapor using a GIS needle in an Ultra-High Vacuum (UHV) environment of $10^{-6}$ mbar. The water is supplied in the form of Magnesium Sulfate Heptahydrate ($\text{MgSO}_{\text{4}}\cdot\text{7}\text{H}_{\text{2}}\text{O}$), which -upon heating- releases water in gaseous form. Water vapor is released locally above the sample using a needle that is inserted a few $\mu$m above the sample surface. The working principle that governs the gas-assisted EBIE method, as presented in Fig.\ref{Exp_setup_HR}, is based on the dissociation of water vapor molecules due to impact ionization from the electrons. The resulting ions then create volatile compounds with the carbon atoms of the diamond surface (volatilization), which are then removed from the sample and eventually pumped out of the chamber\cite{Taniguchi1997}. Subsequently, as described in the second part of the results section, the utilization of the focused ion beam is necessary in order to mill a rectangular area across each trench, to obtain the corresponding cross section and measure the depth of each trench for the calculation of the aspect ratio. For this step, the stage is tilted by 52 degrees. The focused ion beam consists of $Ga^{+}$ ions derived from a Liquid Metal Ion Source (LMIS) source.

\section{Results}

\subsection{High resolution parametric study}

\paragraph{}
In order to achieve the highest resolution of 10 nm and the aspect ratio of 4.5  the influence of the e-beam acceleration voltage and current on the etching process was investigated. It was found that the hydrogen-termination of the diamond samples is crucial for our study because it provides sufficient surface conductivity. Due to the two-dimensional hole gas (2DHG) that is established on the surface of the sample\cite{Maier2000}, charging effects can be avoided and thus, high-resolution imaging and patterning can be achieved.

\begin{figure}[h!]
	\centering
	\includegraphics[width=\linewidth]{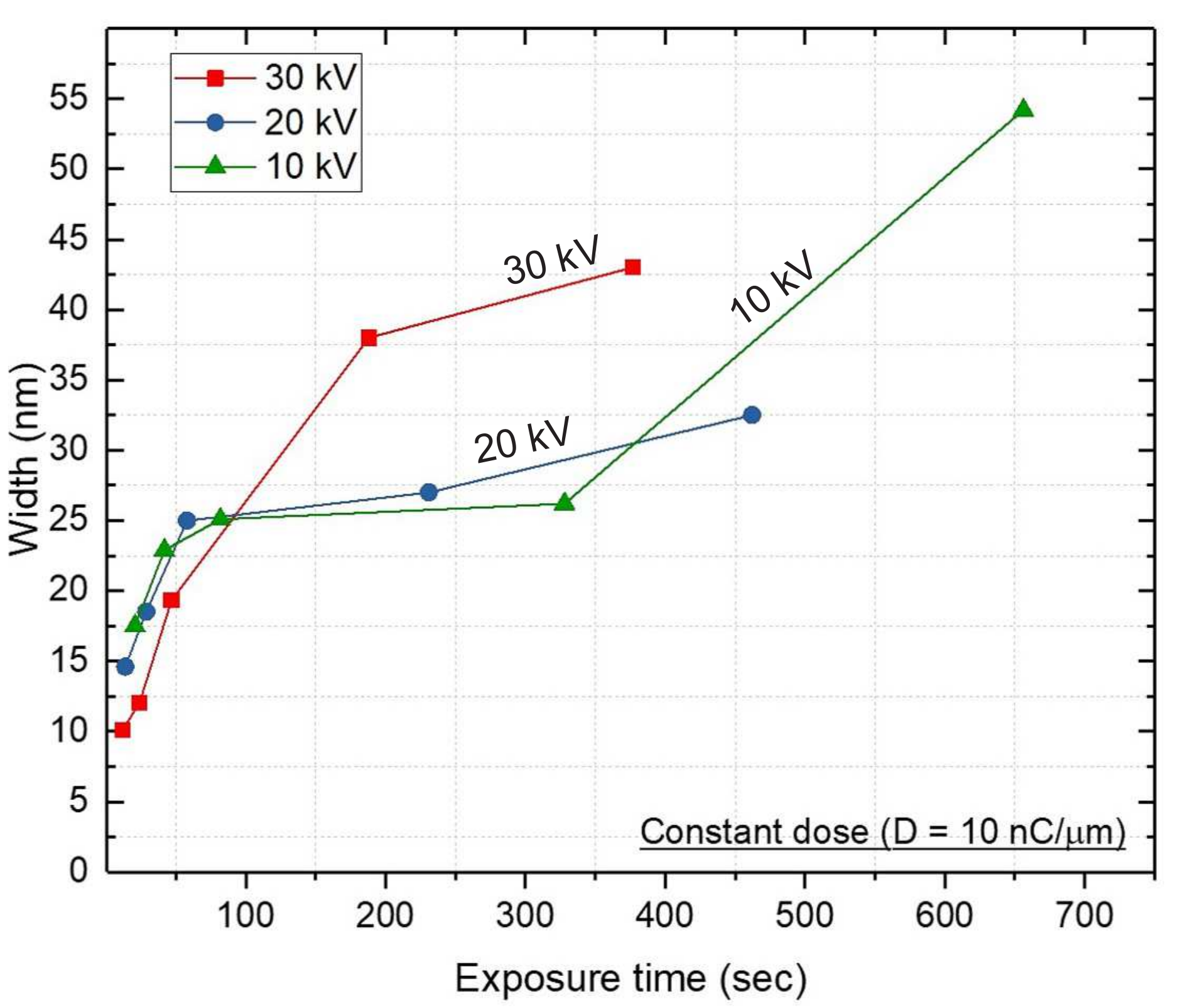}
	\caption{Width of etched trenches for the same dose of approximately 10 nC/$\mu$m and different electron acceleration voltage and beam current. The current ranged from 43 pA to 0.69 nA and the acceleration voltages were 10, 20 and 30 kV. The trench widths range from 54 nm and go down to 10 nm, which is the highest resolution obtained for electron-beam-induced etching of bulk diamond.}
	\label{HR}
\end{figure}

 The single crystalline diamond sample was mounted on the sample holder, inserted in the FIB setup chamber and the system was pumped down to UHV region of $10^{-6}$ mbar. For the trench series, the Ultra High Resolution mode (Immersion) was implemented with the Through The Lens (TTL) detector. Using this mode, trenches of 1 $\mu$m length and only a few nm width were patterned into the diamond surface with varying electron beam current and acceleration voltage and the corresponding trench width was measured in each case. The dose of the e-beam was kept constant for every patterned trench in order to compare the effect of different SEM parameters on the etching process. Note that under constant dose the variation of the beam current, lead to different exposure times to the water vapor and the e-beam. 
 
  As shown in Fig.\ref{HR}, an increase of the current of the electron beam corresponds to reduction of the exposure and etching time, which results to better resolution of the patterned trenches. This indicates that the resolution is given by the temporal stability of the setup. The lowest acceleration voltage of 10 kV and a small current value of 43 pA for 11 minutes, yield a resolution of 54 nm, while increased current and voltage to 0.69 nA and 30 kV for 12 seconds respectively yields trenches of only 10 nm. For short exposure times, i.e. below 60 seconds, the acceleration voltage of 30 kV provides better resolution compared to 20 and 10 kV. However, for longer exposure times, 30 kV of acceleration voltage lead to broader trenches than the 10 and 20 kV. 
  
\subsection{Aspect ratio}

\paragraph{}
In the second part of this study, taking into account the results presented in the previous section and using the beam current that resulted in the best resolution trenches (0.69 nA), a second series of trenches was patterned. The current of the electron beam was kept at 0.69 nA, while the exposure time was increasing from 2 minutes to 5, 15, 30 and 45 minutes and the e-beam dose was calculated respectively. The linear distribution of the e-beam for the aforementioned exposure times correspond to 0.08, 0.2, 0.62, 1.24 and 1.86 $\mu$C/${\mu}$m. The increase of the dose resulted in deeper trenches, however, with increased widths. This effect was investigated by calculating the aspect ratio of each trench.

\begin{figure}[h]
	\centering
	\includegraphics[width=\linewidth]{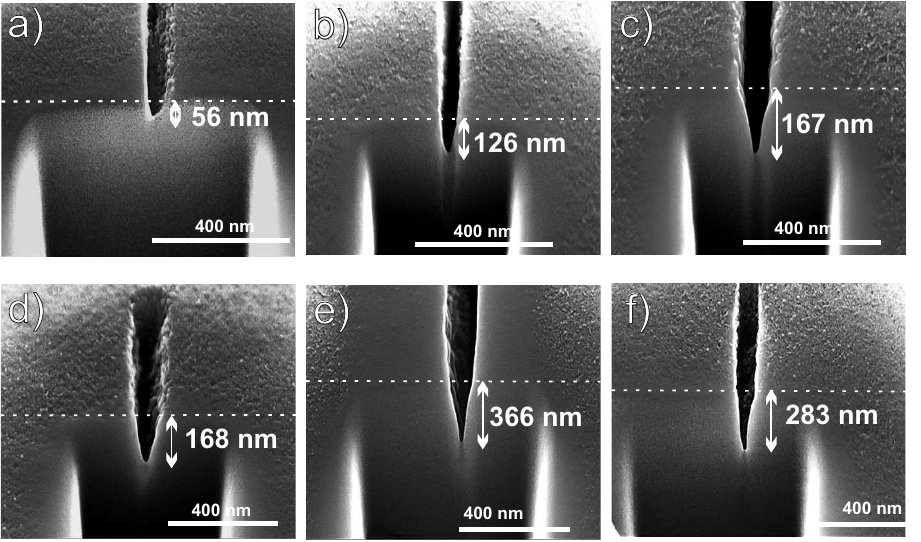}
	\caption{SEM images of trenches patterned with EBIE along with their cross section obtained via FIB milling. a) Trench patterned with 30 kV acceleration voltage and 0.69 nA for 5 minutes exposure time. b) Trench patterned with 20 kV acceleration voltage and 0.69 nA for 15 min exposure time c) Trench patterned with 10 kV acceleration voltage and 0.69 nA for 30 min exposure time. d) Trench patterned with 10 kV acceleration voltage and 0.69 nA for 5 min exposure time. e) Trench patterned with 20 kV acceleration voltage and 0.69 nA for 45 min exposure time. f) Trench patterned with 30 kV acceleration voltage and 0.69 nA for 15 min exposure time.}
	\label{cross_section}
\end{figure}

   \begin{figure}[h!]
	\centering
	\includegraphics[width=\linewidth]{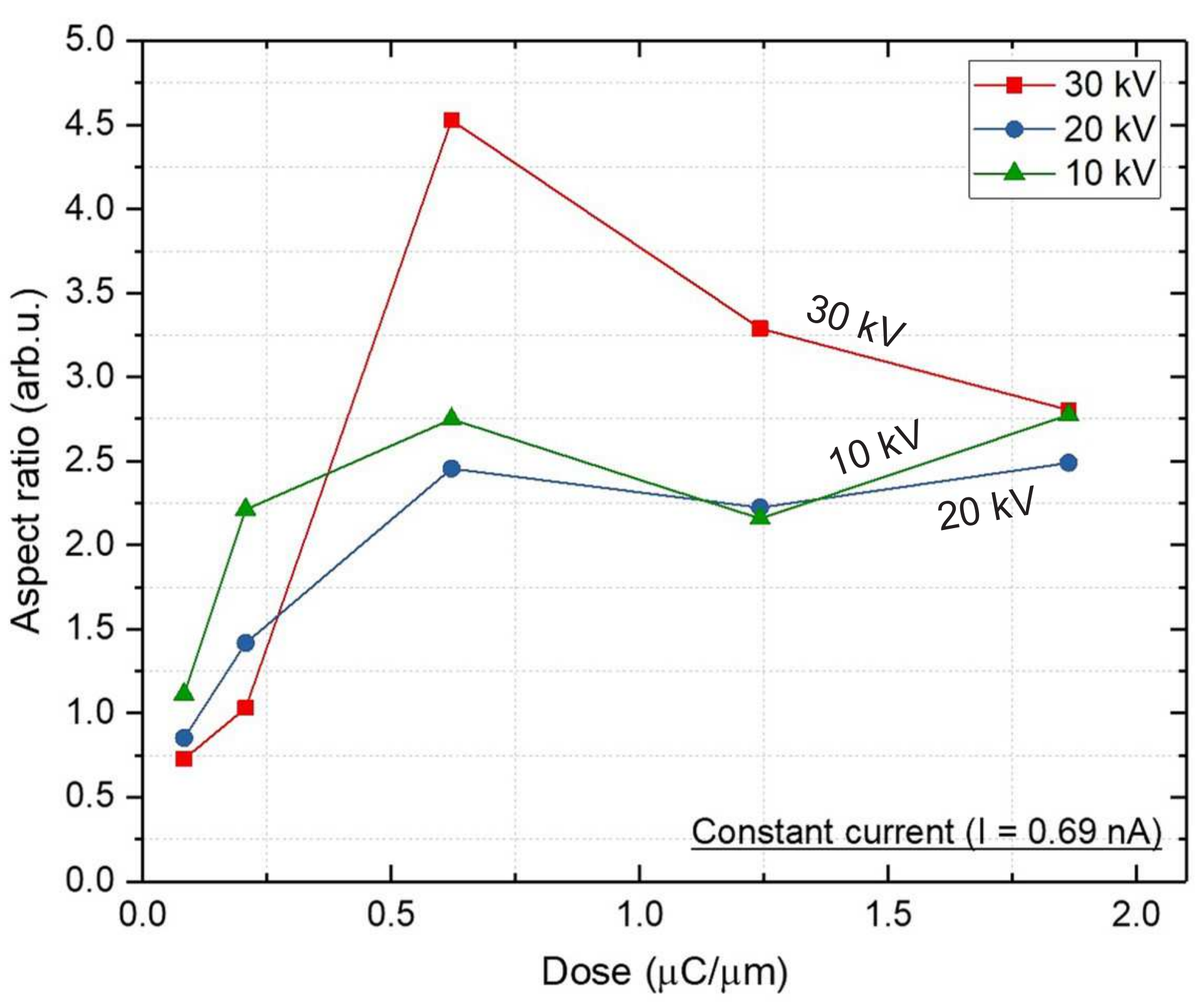}
	\caption{Aspect ratio as a function of electron-beam dose for trenches etched with acceleration voltages 10, 20 and 30 kV and constant current of 0.69 nA. The aspect ratio values, range from 1 to 4.5. From 0 to 0.6 $\mu$C/$\mu$m, the aspect ratio increases, while above this value it seems to be fluctuating between 2 and 4.5.}
	\label{aspect_ratio_graph}
\end{figure}

 For a stable current value of 0.69 nA and acceleration voltages of 10, 20 and 30 kV, 15 trenches were patterned in total for 2,5, 15, 30 and 45 minutes of exposure time. The trench widths were measured and then the stage was tilted to \ang{52} in order to employ the FIB to mill rectangular holes of 600x600 nm to 1 $\mu$m depth and measure the cross section of the trenches, see Fig.\ref{cross_section}. Using the measured widths and cross sections, the aspect ratios for the different acceleration voltages were calculated and are presented in Fig.\ref{aspect_ratio_graph}. The obtained aspect ratios are calculated to be between 1:1 and 1:5.
 
  For small exposure times (5 and 15 min) the aspect ratio is around 1:1, while with increasing the exposure time, the depth of the trenches is increased up to 385 nm. However, by increasing the exposure time, a broadening is also observed in most trenches. This effect is attributed to increased number of passes in combination with a minor sample drift and external vibrations. The broadening resulted in increased widths and an asymmetry of trenches that lead to variations of the aspect ratios despite the increased depths. It is noteworthy, that the broadening of the trenches is not completely avoidable even if the drift is reduced. A possible reason for this is that the water molecules that are being adsorbed from a larger area around the etched trench, can also be dissociated by the secondary electrons created under the diamond surface. This could explain the reversed conical shape of the trenches that can be seen at Fig.\ref{cross_section}. The latter effect of the cone-shaped broadened trenches should be also attributed to the reduced diffusion of the water molecules in the trench during etching process. Under the conditions maintained during etching ($<10^{-4}$), the fluid dynamics are described by the free molecular flow. As a consequence, the random movement of the water molecules inhibits their further diffusion in the etched trench, thus reducing the thickness gradually.

\section{Conclusion}
\paragraph{}
In conclusion, we have studied in depth the EBIE method on diamond. By systematically varying the etching parameters, we were able to obtain an exceptionally high resolution of only 10 nm for patterned trenches. The influence of the different parameters on the etching process was investigated in depth and defined accurately. Furthermore, the aspect ratio of the patterned trenches was calculated, in order to obtain an estimate about the removed material. Apart from the optimum choice of the etching parameters, the hydrogen termination of the diamond samples was of great importance to reach high resolution imaging and patterning, because it provided the necessary surface conductivity that minimized the charging effects. The reported results can be improved even further by reducing the number of e-beam passes over the patterned area, which will limit the observed broadening. Our results open up new ways of easier processing of diamond towards the fabrication of high
precision devices at the nanometer range that exploit the unique properties of this special material.

\section{Acknowledgments}

This work was financially supported by the International Max Planck Research School for Interface Controlled Materials for Energy Conversion (IMPRS-SurMat).

\section{References}

\bibliography{bib2}

\end{document}